\documentstyle[a4,11pt,epsfig,twoside,cite]{article}

\expandafter\ifx\csname mathrm\endcsname\relax\def\mathrm#1{{\rm #1}}\fi

\def\bfi{\begin{figure}}
\def\efi{\end{figure}}

\newcommand{\lsim}
{\mathrel{\raisebox{-.3em}{$\stackrel{\displaystyle <}{\sim}$}}}
\newcommand{\gsim}
{\mathrel{\raisebox{-.3em}{$\stackrel{\displaystyle >}{\sim}$}}}
\def\asymp#1%
{\mathrel{\raisebox{-.4em}{$\widetilde{\scriptstyle #1}$}}}

\def\Nequal#1%
{\mathrel{\raisebox{-.5em}{$\stackrel{=}{\scriptstyle\rm#1}$}}}

\def\reffi#1{\mbox{Figure~\ref{#1}}}

\def\refta#1{\mbox{Table~\ref{#1}}}

\def\citere#1{\mbox{Ref.~\cite{#1}}}

\newcommand{\GeV}{\unskip\,\mathrm{GeV}}
\newcommand{\MeV}{\unskip\,\mathrm{MeV}}

\def\mathswitchr#1{\relax\ifmmode{\mathrm{#1}}\else$\mathrm{#1}$\fi}
\newcommand{\Pf}{\mathswitch  f}

\newcommand{\PW}{\mathswitchr W}

\newcommand{\PZ}{\mathswitchr Z}

\newcommand{\Pe}{\mathswitchr e}

\newcommand{\Pd}{\mathswitchr d}
\newcommand{\Pdbar}{\bar{\mathswitchr d}}
\newcommand{\Pu}{\mathswitchr u}

\newcommand{\Pep}{\mathswitchr {e^+}}
\newcommand{\Pem}{\mathswitchr {e^-}}

\def\mathswitch#1{\relax\ifmmode#1\else$#1$\fi}

\newcommand{\MW}{\mathswitch {M_\PW}}

\newcommand{\Me}{\mathswitch {m_\Pe}}

\def\draftdate{\relax}
\def\mda{\relax}
\def\mua{\relax}
\def\mla{\relax}
\def\draft{
\def\thtystars{******************************}
\def\sixtystars{\thtystars\thtystars}
\typeout{}
\typeout{\sixtystars**}
\typeout{* Draft mode!
         For final version remove \protect\draft\space in source file *}
\typeout{\sixtystars**}
\typeout{}
\def\draftdate{\today}
\def\mua{\marginpar[\boldmath\hfil$\uparrow$]%
                   {\boldmath$\uparrow$\hfil}%
                    \typeout{marginpar: $\uparrow$}\ignorespaces}
\def\mda{\marginpar[\boldmath\hfil$\downarrow$]%
                   {\boldmath$\downarrow$\hfil}%
                    \typeout{marginpar: $\downarrow$}\ignorespaces}
\def\mla{\marginpar[\boldmath\hfil$\rightarrow$]%
                   {\boldmath$\leftarrow $\hfil}%
                    \typeout{marginpar: $\leftrightarrow$}\ignorespaces}
\def\Mua{\marginpar[\boldmath\hfil$\Uparrow$]%
                   {\boldmath$\Uparrow$\hfil}%
                    \typeout{marginpar: $\uparrow$}\ignorespaces}
\def\Mda{\marginpar[\boldmath\hfil$\Downarrow$]%
                   {\boldmath$\Downarrow$\hfil}%
                    \typeout{marginpar: $\downarrow$}\ignorespaces}
\def\Mla{\marginpar[\boldmath\hfil$\Rightarrow$]%
                   {\boldmath$\Leftarrow $\hfil}%
                    \typeout{marginpar: $\leftrightarrow$}\ignorespaces}
\overfullrule 5pt
\oddsidemargin -15mm
\marginparwidth 29mm
}

\def\stars{\strut\leaders\hbox{*}\hfill\strut}
\def\starline{\hfil\strut\hfil\hbox to \textwidth {\stars}\hfil}

\begin{document}

\thispagestyle{empty}
\def\thefootnote{\fnsymbol{footnote}}
\setcounter{footnote}{1}
\null
\draftdate\hfill KA-TP-17-2002 \\
\strut\hfill MPI-PhT/2002-57\\
\strut\hfill PSI-PR-02-13\\
\strut\hfill UB-HET-02-07\\
\strut\hfill hep-ph/0210169 
\vfill
\vspace{.5cm}
\begin{center}
{\large \bf\boldmath{CHALLENGES IN W-PAIR PRODUCTION}%
\footnote{To appear in the proceedings of the {\it International Workshop 
on Linear Colliders}, August 26--30, 2002, Jeju Island, Korea.}
\par} \vskip 1em
\vspace{.1cm}
{\large
{\sc A.\ Denner$^1$, S.\ Dittmaier$^2$,
M. Roth$^{3}$ and D.\ Wackeroth$^4$} } 
\\[.5cm]
$^1$ {\it Paul Scherrer Institut\\
CH-5232 Villigen PSI, Switzerland} 
\\[0.3cm]
$^2$ {\it Max-Planck-Institut f\"ur Physik (Werner-Heisenberg-Institut) \\
D-80805 M\"unchen, Germany}
\\[0.3cm]
$^3$ {\it Institut f\"ur Theoretische Physik, Universit\"at Karlsruhe \\
D-76131 Karlsruhe, Germany}
\\[0.3cm]
$^4$ {\it Department of Physics, SUNY at Buffalo, NY 14260, USA} 
\par 
\end{center}\par
\vskip 1.0cm {\bf Abstract:} \par 
The investigation of W-pair production offers unique precision tests
of the electroweak theory at future $\Pep\Pem$ colliders, including
precise determinations of cross sections, the W-boson mass, and 
gauge-boson self-couplings. The 
state-of-the-art and future
requirements in the theoretical prediction for 
the reaction $\Pep\Pem\to\PW\PW\to 4f(+\gamma)$
are briefly reviewed.
\par
\vskip .7cm
\noindent
October 2002
\null
\setcounter{page}{0}

\clearpage

\setcounter{page}{1}
\title{CHALLENGES IN W-PAIR PRODUCTION}
\author{A.~DENNER$^1$, S.~DITTMAIER$^2$%
        \thanks{e-mail address: dittmair@mppmu.mpg.de},
        {} M.~ROTH$^3$ and D.~WACKEROTH$^4$
\\
\\
        $^1$ {\it Paul Scherrer Institut, CH-5232 Villigen PSI, Switzerland}
\\
\\
        $^2$ {\it Max-Planck-Institut f\"ur Physik (Werner-Heisenberg-Institut)}
\\
             {\it D-80805 M\"unchen, Germany}
\\
\\
        $^3$ {\it Institut f\"ur Theoretische Physik, Universit\"at Karlsruhe}
\\
             {\it D-76131 Karlsruhe, Germany}
\\
\\
        $^4$ {\it Department of Physics, SUNY at Buffalo, NY 14260, USA} 
}
\date{}
\maketitle
\begin{abstract}
The investigation of W-pair production offers unique precision tests
of the electroweak theory at future $\Pep\Pem$ colliders, including
precise determinations of cross sections, the W-boson mass, and 
gauge-boson self-couplings. The 
state-of-the-art and future
requirements in the theoretical prediction for 
the reaction $\Pep\Pem\to\PW\PW\to 4f(+\gamma)$
are briefly reviewed.
\end{abstract}

\section{Introduction}

At LEP2, W-pair-mediated four-fermion production was 
experimentally explored with quite high precision \cite{lep2}.
The total W-pair cross section was measured from threshold
up to a centre-of-mass (CM) energy of $209\GeV$; combining the
cross-section measurements a precision of $\sim 1\%$ was reached. The
W-boson mass $\MW$ was determined from the threshold cross section
with an error of $\sim 200\MeV$ and by 
reconstructing the $\PW$ bosons from their decay products
within $\sim 40\MeV$, where a 
further reduction of the error down to $\sim 35\MeV$ is expected.
Deviations from the Standard Model (SM) triple gauge-boson couplings,
usually quantified in the parameters $\Delta g^\PZ_1$,
$\Delta\kappa_\gamma$, and $\lambda_\gamma$, were constrained within a
few per cent.  At a future $\Pe^+ \Pe^-$ linear collider
\cite{Accomando:1998wt,Aguilar-Saavedra:2001rg,Abe:2001wn,Abe:2001gc}, 
the accuracy of the cross-section measurement will be at the per-mille
level, and the precision of 
the $\PW$-mass determination is expected to be $15 \MeV$ by 
direct reconstruction and about $6 \MeV$ from a threshold scan of the total
W-pair-production cross section.

The precision reached at LEP2 triggered considerable theoretical
progress in the past years, as it is reviewed in
Refs.~\cite{Beenakker:1996kt,Grunewald:2000ju}.  In the present
calculations, the $\PW$ bosons are treated as resonances in the full
4-fermion processes, $\Pe^+ \Pe^- \to 4 \Pf\, (+\,\gamma)$, and
radiative corrections (RC) are taken into account in a proper way.
The RC can be split into universal and non-universal corrections.  The
former comprise leading-logarithmic 
(LL) corrections from
initial-state radiation (ISR), higher-order corrections included by
using 
appropriate effective couplings, and the Coulomb singularity.  The
remaining corrections are called non-universal since they depend on
the process under investigation.  
Since the full ${\cal O}(\alpha)$
corrections to the $4f$ processes are not necessary to match the
accuracy of LEP2, 
it is sufficient to take only those corrections
into account that are enhanced by two resonant $\PW$~bosons.  The
leading term of an expansion about the two $\PW$ poles provides the
so-called double-pole approximation (DPA) \cite{DPA}. Different
versions of such a pole approximation
have been used in the literature
\cite{DPAversions,Denner:2000bj,Jadach:2000kw}.  Although several
Monte Carlo programs exist that include universal corrections, only
two event generators, {\sc YFSWW}
\cite{Jadach:2000kw,Jadach:1995sp,Jadach:2001uu} 
and {\sc RacoonWW}
\cite{Denner:2000bj,Denner:1999gp,Denner:2000kn,Denner:2001zp,Denner:2002cg},
include non-universal corrections.  While the DPA approach is
sufficient for the LEP2 accuracy
\cite{Grunewald:2000ju,Jadach:2001cz,Bruneliere:2002df}, the extremely
high experimental precision at a future linear collider is a great
challenge for future theoretical predictions. 
Moreover, the
DPA is not reliable near the $\PW$-pair threshold. In the following,
the necessary theoretical improvements will be discussed in some
detail.

\section{Total Cross Section}

The W-pair cross-section measurements at LEP2 have tested the SM
predictions at the per-cent level in the energy range between 170 and
$209\GeV$, thereby rendering non-leading (NL) electroweak corrections
of non-universal origin, which are about 2\%, experimentally
significant.  The presently available calculations, provided by
{\sc YFSWW} and {\sc RacoonWW}, which are both based on a DPA, involve a 
theoretical uncertainty (TU) of about $\sim 0.5\%$ in the range
between 170 and $500\GeV$
\cite{Grunewald:2000ju,Denner:2000bj,Jadach:2000kw}.  This estimate 
emerges from a detailed comparison between the two programs and
from investigations of intrinsic uncertainties in the DPA versions.
The l.h.s.\ of \reffi{fig:cs} compares the full DPA RC with two
improved Born approximations (IBA) that are based on universal
corrections only, which illustrates that non-universal RC become more
and more important at higher energies.
\bfi
\setlength{\unitlength}{1cm}
\centerline{
\begin{picture}(7.3,6.2)
\put(-2.8,-14.4){\includegraphics{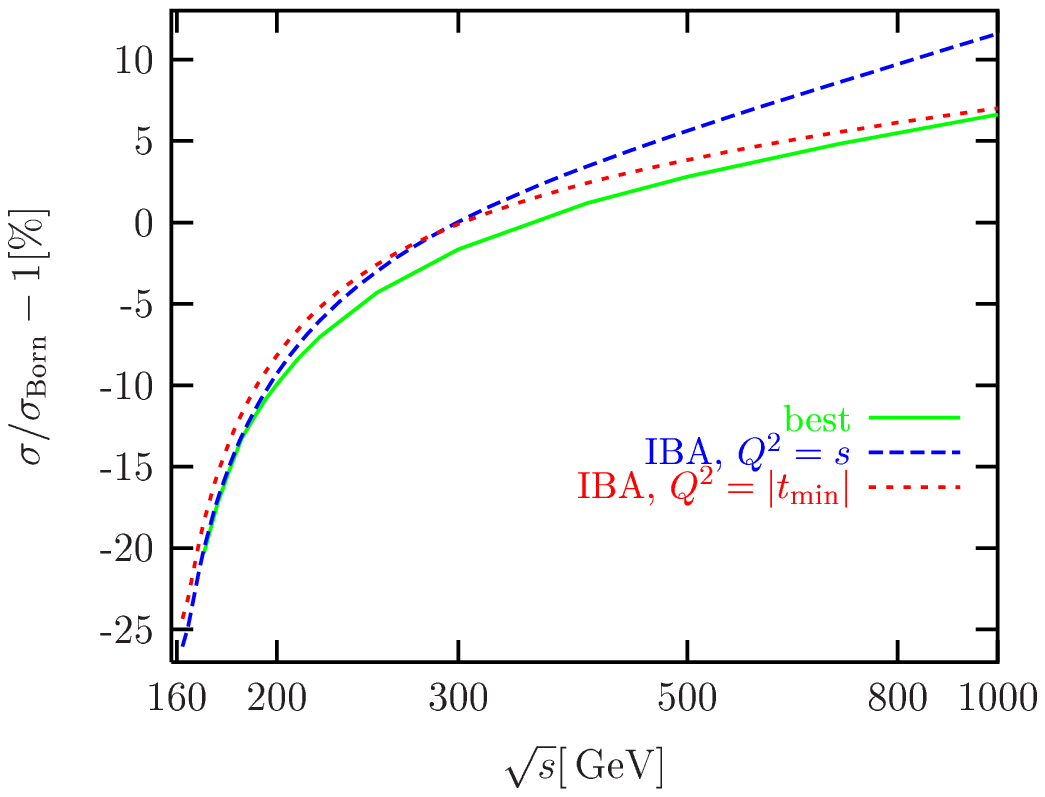}}
\end{picture}
\begin{picture}(6,6.2)
\put(0,0){\epsfig{file=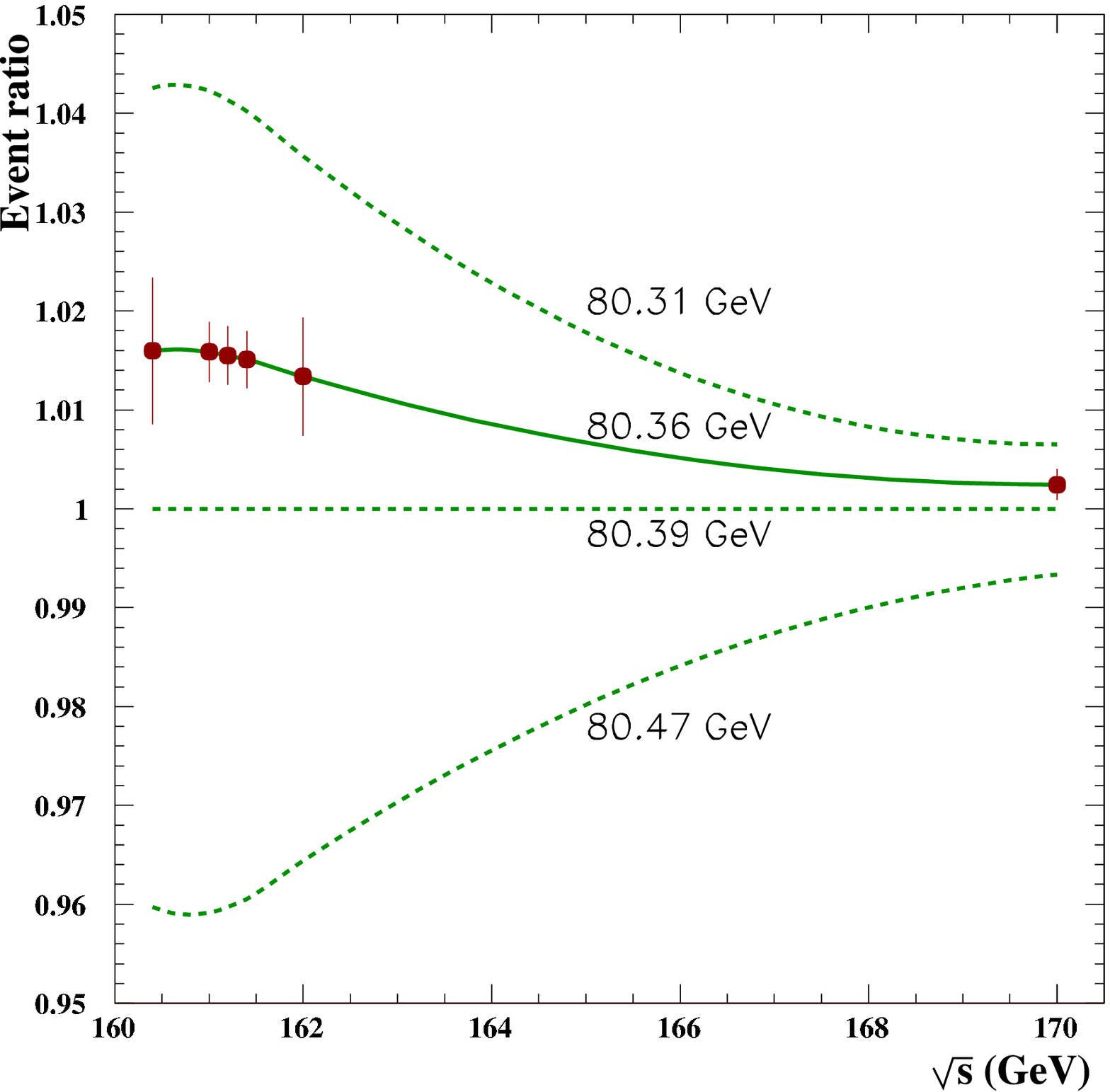, scale=.35}}
\put(0.645,4.42){{\line(1,0){5.59}}}
\put(0.645,2.18) {{\line(1,0){5.59}}}
\end{picture}} 
\caption{L.h.s.: 
  Relative corrections to the total cross section for the process
  $\Pep\Pem\to\Pu\bar\Pd\mu^-\bar\nu_\mu$ based on the full DPA of
  {\sc RacoonWW} (``best'') and on two IBA versions (taken from
  \citere{Denner:2001zp}); r.h.s.: Hypothetical data points at a
  future LC in comparison with cross-section predictions for various
  values of $\MW$ (taken from \citere{Aguilar-Saavedra:2001rg}), where
  the lines at $\pm 2\%$ indicate the TU from neglecting NL
  corrections.}
\label{fig:cs}
\efi
Measurements at future LCs, which will be precise within a few
per mille, require a TU of 0.1\% or better.  In order to illustrate
the consequences of this requirement on theoretical predictions, 
in \refta{tab:csTU}
we collect some estimates of corrections that are neglected in present
calculations.
\begin{table}
\centerline{
\begin{tabular}{|l|c|}
\hline
Neglected effect & Estimate for relative numerical impact \\
\hline \hline
RC to background diagrams & 
$(\alpha/\pi)\times(\Gamma_\PW/\MW)\times\mathrm{const} \;\sim\; 0.1\%$ \\
\hline
Scale in coupling of NL RC & 
$(\Delta\alpha/\alpha)\times\mathrm{NL} \;\sim\; 6\%\times 2\%
\;\sim\; 0.1\%$ \\
\hline
Squared NL corrections & $(\mathrm{NL})^2 \;\sim\; 0.04\%$ \\
\hline
Interference of NL and ISR & 
${\mathrm{NL}}\times{\mathrm{ISR}} \;\sim\; 2\%\times(\alpha/\pi)\ln({\mathit{s}}/{\mathit{\Me}}^2)
\;\sim\; 0.1\%$ \\
\hline
\end{tabular} }
\caption{Estimates of some presently missing RC to the total W-pair cross 
section at $\sqrt{s}\sim 200\GeV$}
\label{tab:csTU}
\end{table}
The table shows that there is a variety of neglected terms potentially
of the order of 0.1\%, and that an improvement on the TU to this level
requires a full ${\cal O}(\alpha)$ calculation of the processes 
$\Pep\Pem\to 4f$ and a proper inclusion of the most important two-loop
effects. 
While the virtual one-loop RC to $\Pep\Pem\to 4f$ are not known yet,
the real ${\cal O}(\alpha)$ RC, which are induced by the processes
$\Pep\Pem\to 4f+\gamma$, are available
\cite{Denner:1999gp,Papadopoulos:2000tt}. However, at ${\cal
  O}(\alpha^2)$, real photonic corrections beyond the 
LL approximation and the effects of collinear emission of 
$f\bar f$ pairs must be included.
\looseness -1

The necessity of the full treatment of $\Pep\Pem\to 4f$ at one loop
becomes even more obvious near the W-pair threshold 
($\sqrt{s}\lsim 170\GeV$) where the TU of the DPA approach runs out
of control
because of the increasing relative importance of the non-resonant contribution.
Therefore, at LEP2 an IBA was confronted with the cross section
measured at $\sqrt{s}=161\GeV$, since the experimental error of 12\%\ was
much larger than the IBA uncertainty of about 2\%.
The r.h.s.\ of \reffi{fig:cs} shows the possible result of a threshold
scan at a future LC running with high luminosity and the sensitivity 
of the cross section to the W-boson mass. Without reducing the TU to
the level of 
a few 0.1\%\ the aimed precision of $6\MeV$ in the $\MW$
determination will be impossible.
\looseness -1

At high scattering energies, $\sqrt{s}\gg\MW$, (and fixed angles) the
RC are dominated by electroweak logarithms of the form
$[\alpha\ln^2(s/\MW^2)]^n$, known as Sudakov logarithms, and single
logarithms like $[\alpha\ln(s/\MW^2)]^n$.  While these terms are
implicitly contained in the present DPA approaches at the one-loop
level, the higher-order logarithms, $n\ge2$, are not yet included in
existing generators.  These missing terms are potentially numerically
relevant for $\sqrt{s}\gsim500\GeV$. The existing efforts in the
calculation of these logarithms in virtual corrections are reviewed in
\citere{sudakovlogs}. In addition also the corresponding logarithms
from real corrections and enhanced logarithms resulting from small
scattering angles (Regge limit) have to be investigated.

\section{Invariant-Mass Distributions and W-Boson Mass}

The invariant-mass distributions of the W~bosons are the central
observables in the $\MW$ determination from the direct reconstruction
of the W~bosons from their decay products. While the overall scale of
the distributions more or less reflects the situation of the total
cross section, the shapes and peak positions of the Breit--Wigner-type
resonances are sensitive to $\MW$.  Based on a comparison between {\sc
  YFSWW} and {\sc RacoonWW}, the present TU induced by missing RC was
estimated to 
$\lsim 1\%$ in the invariant-mass distributions \cite{Grunewald:2000ju}
and to $\sim 5\MeV$ in the reconstructed W-boson mass
\cite{Jadach:2001cz}, which is small
compared to the aimed precision of 
$35\MeV$ at LEP2. Note that the TU
is also smaller than the expected accuracy of $15\MeV$ at a LC, but a
further reduction of the TU would certainly be welcome. This
improvement would, however, also require a much better understanding
of QCD corrections that are connected with the W~decays and a proper
matching between parton-level calculations and hadronization
procedures.

\section{Angular Distributions and Anomalous Couplings}

A proper way to study possible deviations from the SM 
triple gauge-boson couplings
is the analysis of angular distributions, where the W-production angle
plays the most important role.
In \reffi{fig:atgc} we compare the influence of anomalous charged 
triple gauge-boson couplings with the effect of the non-universal
corrections for 
the process $\Pe^+ \Pe^- \to \Pu \Pdbar \mu^-
\bar\nu_\mu$ at the 
CM energy of $\sqrt{s}=200\GeV$.
\bfi
\setlength{\unitlength}{1cm}
\begin{picture}(6.5,7.6)
\put(-2.0,-15.5){\includegraphics{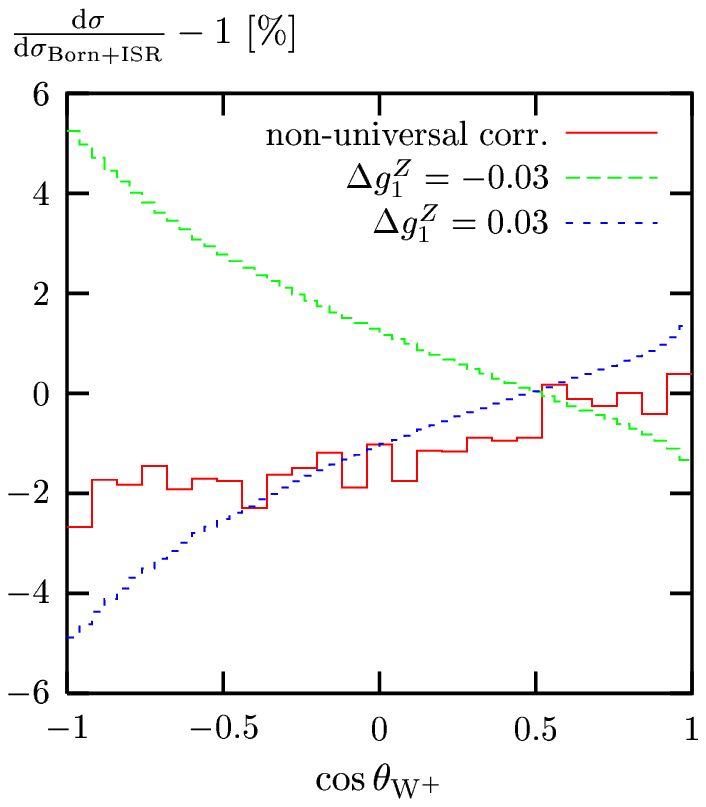}}
\end{picture}
\setlength{\unitlength}{1cm}
\begin{picture}(6.5,7.6)
\put(-2.0,-15.5){\includegraphics{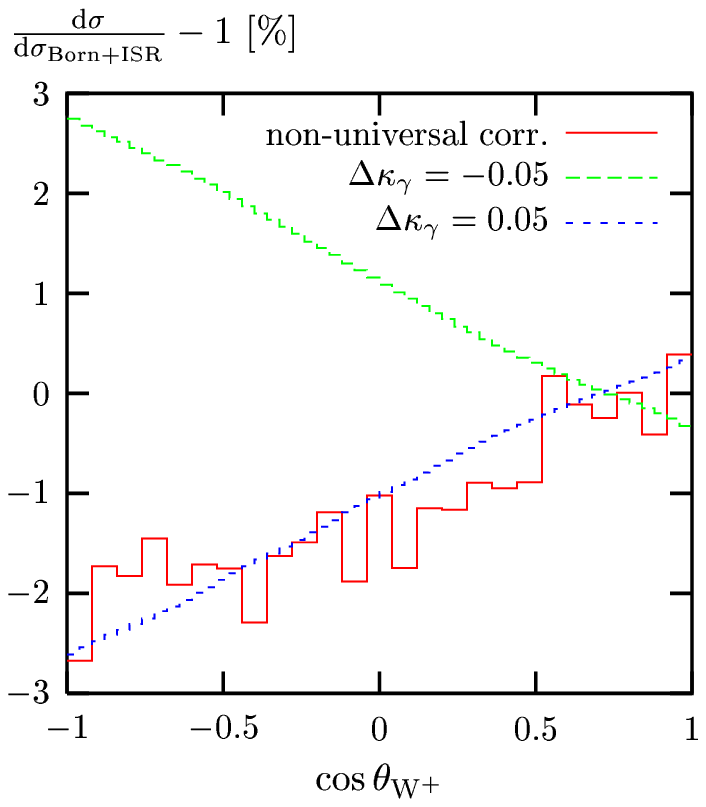}}
\end{picture} 
\\[.5em]
\begin{picture}(13,7.6)
\put(-2.0,-15.5){\includegraphics{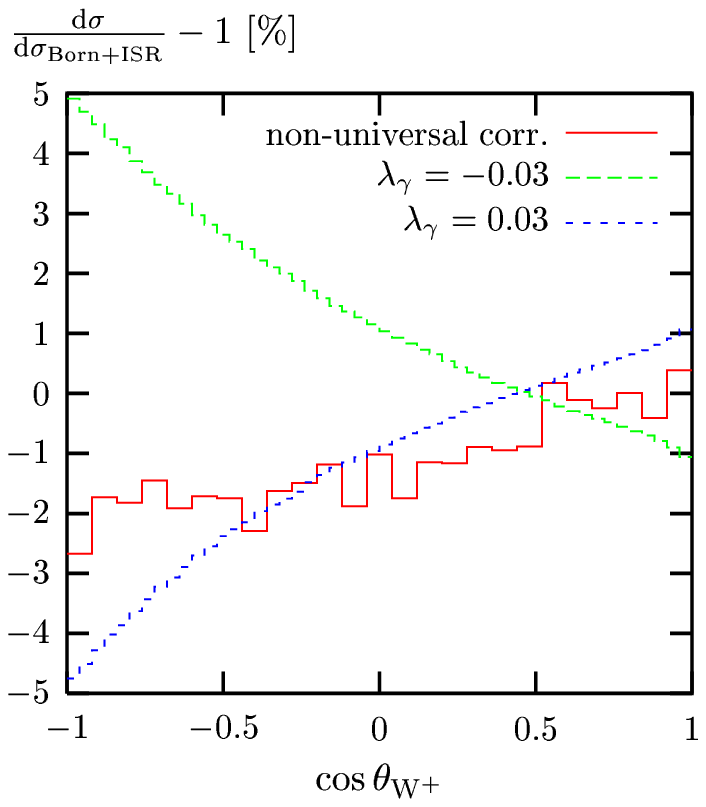}}
\put(8,5.4){\Large\textsf{RacoonWW}}
\put(8,4.4){$\Pep\Pem\to\Pu\bar\Pd\mu^-\bar\nu_\mu$}
\put(8,3.6){$\sqrt{s}=200\GeV$}
\put(8,2.8){$\lambda_\PZ=\lambda_\gamma$}
\put(8,2){$\Delta\kappa_\PZ=\Delta g^\PZ_1+\tan^2\theta_\PW\Delta\,\kappa_\gamma$}
\end{picture} 
\caption{Influence of anomalous triple gauge-boson couplings and 
  non-universal corrections in the $\PW^+$-production-angle
  distribution (taken from \citere{Denner:2001bd})}
\label{fig:atgc}
\efi
Following a convention widely used in the LEP2 data analysis, we
consider only the coupling constants $g^Z_1$, $\kappa_\gamma$, and
$\lambda_\gamma$, where $\Delta$ indicates the deviation from the
SM values $g_1^Z=\kappa_\gamma=1$ and
$\lambda_\gamma=0$.  In the figures all numbers are normalized to the
tree-level cross section including higher-order LL ISR.  The relative
deviations for different values of the anomalous couplings are
compared with the 
predictions including non-universal
${\cal O}(\alpha)$ corrections instead of anomalous couplings.  The
labels indicate the values of the corresponding anomalous coupling
constants, which are chosen to be of the order of the actual accuracy
achieved by the LEP experiments, i.e.\ of the order of a few per cent.
The comparison shows clearly that the non-universal corrections are of
the same size as the possible 
contributions from anomalous couplings and, thus,
had to be taken into account in the determination of limits on 
anomalous couplings at LEP2. 
The angular distributions obtained with {\sc YFSWW} and {\sc RacoonWW} 
differ by about $\lsim 1\%$ \cite{Grunewald:2000ju}.
A detailed analysis \cite{Bruneliere:2002df} showed that
this TU, for instance, leads to an error of $5\times 10^{-3}$ 
in the parameter $\lambda_\gamma$,
which is sufficient for
LEP2.  At a future LC, however, the sensitivity to the parameters
$g^Z_1$, $\kappa_\gamma$, and $\lambda_\gamma$ will be of the order of
$10^{-3}$, so that drastic improvements in the predictions are
necessary in order to match this precision. Again the inclusion
of the full ${\cal O}(\alpha)$ corrections
to $\Pep\Pem\to 4f$ and improvements by 
higher-order corrections are indispensable.
\looseness -1

\end{document}